\documentclass{aa}

\usepackage{graphicx}
\usepackage{txfonts}
\usepackage{dcolumn}
\usepackage{bm}
\usepackage{xspace}
\usepackage{xcolor}
\usepackage{hyperref}
\hypersetup{
  colorlinks,
  citecolor=blue,
  linkcolor=blue,
  urlcolor=blue}

\usepackage[mathlines]{lineno}
\usepackage{physics}
\usepackage{booktabs}
\usepackage[normalem]{ulem}

\newcommand{\AREPO}{\textsc{arepo}\xspace}

\newcommand{\msol}{\,$\textrm{M}_\odot$\xspace}

\newcommand{\E}[1]{\,{\times}\,10^{#1}\xspace}
\newcommand{\eb}{$\textrm{E}_\textrm{B}$\xspace}

\makeatletter
\renewcommand*\aa@pageof{, page \thepage{} of \pageref*{LastPage}}
\makeatother

\begin{document}

\title{Self-consistent MHD simulation of jet launching in a neutron star - white dwarf merger}
\author{
    Javier Mor\'an-Fraile\inst{1}
     \and
    Friedrich K. Röpke\inst{1,2}
    \and
    Rüdiger Pakmor\inst{3}
    \and
    Miguel A. Aloy\inst{4,5}
    \and
    Sebastian T. Ohlmann\inst{6}
    \and
    Fabian R. N. Schneider\inst{1,7}
    \and
    Giovanni Leidi\inst{1}
    }

\institute{
    Heidelberger Institut für Theoretische Studien (HITS),
              Schloss-Wolfsbrunnenweg 35, 69118 Heidelberg, Germany
    \and
    Zentrum f\"ur Astronomie der Universit\"at Heidelberg,
    Institut f\"ur Theoretische Astrophysik, 
    Philosophenweg 12,
    69120 Heidelberg, Germany
    \and
    Max-Planck-Institut f\"ur Astrophysik,
    Karl-Schwarzschild-Str. 1, 85748 Garching,
    Germany
    \and
    Departament d'Astonom\'{\i}a i Astrof\'{\i}sca, Universitat de Val\`encia,
    E-46100 Burjassot (Val\`encia), Spain
    \and
    Observatori Astron\`omic, Universitat de Val\`encia,
    46980 Paterna, Spain
    \and
    Max Planck Computing and Data Facility, 
    Gie{\ss}enbachstra{\ss}e 2, 
    85748 Garching, Germany
    \and
    Zentrum f{\"u}r Astronomie der Universit{\"a}t Heidelberg,
    Astronomisches Rechen-Institut,
    M{\"o}nchhofstr.\ 12-14, 
    69120 Heidelberg, Germany
}

\date{\today}

\abstract{
The merger of a white dwarf (WD) and a neutron star (NS) is a relatively common event that will produce an observable electromagnetic signal. Furthermore, the compactness of these stellar objects makes them an interesting candidate for gravitational wave (GW) astronomy, potentially being in the frequency range of LISA and other missions. To date, three-dimensional simulations of these mergers have not fully modelled the WD disruption, or have used lower resolutions and have not included magnetic fields even though they potentially shape the evolution of the merger remnant. In this work, we simulate the merger of a 1.4\msol NS with a 1\msol carbon oxygen WD in the magnetohydrodynamic moving mesh code \AREPO. We find that the disruption of the WD forms an accretion disk around the NS, and the subsequent accretion by the NS powers the launch of strongly magnetized, mildly relativistic jets perpendicular to the orbital plane. Although the exact properties of the jets could be altered by unresolved physics around the NS, the event could result in a transient with a larger luminosity than kilonovae. We discuss possible connections to fast blue optical transients (FBOTs) and long-duration gamma-ray bursts. We find that the frequency of GWs released during the merger is too high to be detectable by the LISA mission, but suitable for deci-hertz observatories such as LGWA, BBO or DECIGO.
}

\maketitle

\section{Introduction \label{sec:intro}}
Binary systems formed by a neutron star (NS) and a white dwarf (WD) are one of the most abundant double compact object systems in the Galaxy \citep{Nelemans2001}, second only to WD-WD systems. 
If the two compact objects are born with a short orbital separation, the release of energy in the form of gravitational waves (GW) can make the binary system merge on timescales less than the age of the Universe. This process can be accelerated if the binary has undergone a common-envelope (CE) phase, where the drag in the envelope drives the orbital shrinking and may even lead to a prompt merger.
When the orbit has decreased to the point where the WD fills its Roche lobe, a mass transfer episode commences.
The stability of this mass transfer depends primarily on the mass of the WD. For sufficiently low-mass WDs, the mass transfer will be stable, and the system can be observed as an ultra-compact X-ray binary \citep{Nelson1986}.
If the WD mass is above a critical value of $M_{\textrm{WD}} = 0.2$\msol \citep{Bobrick2017}, the mass transfer will be unstable and the system will merge on a dynamical timescale, leading to a variety of electromagnetic transients \citep{Metzger2012, Fernandez2019, Zenati2019,Zenati2020,Bobrick2022, Kaltenborn2022} and a burst of GWs, similar to the ones released by binary black hole (BH) mergers \citep{Abbott+2016} and binary NS mergers \citep{Abbott+2017} that have been detected by the LIGO\footnote{Laser Interferometer Gravitational-Wave Observatory}-Virgo-KAGRA\footnote{Kamioka Gravitational Wave Detector}(LVK) collaboration \citep{Aasi+2015,Acernese+2014,Akutsu2021}, albeit in a different frequency range.

The GWs released by these systems prior to the merger make these binaries a likely source \citep{VanZeist2023, Abdusalam2020} for the LISA\footnote{Laser Interferometer Space Antenna} mission \citep{Robson2019}. The GWs released by the merger itself will have higher frequency and will be better captured by proposed deci-hertz observatories such as space-based DECIGO\footnote{DECI-hertz Interferometer Gravitational wave Observatory}\citep{Seto2001} and BBO\footnote{Big Bang Observer}\citep{Phinney2003} or Moon-based LGWA\footnote{Lunar Gravitational-Wave Antenna}\citep{Harms2020}. The post-merger signal which would originate as accretion onto the NS may excite oscillation modes of the NS \citep{Schutz2008} that would be in the frequency range of the LVK network \citep{Aasi+2015,Acernese+2014,Akutsu2021}.

The electromagnetic signal released during these mergers has been proposed to explain several fast-evolving transients such as AT2018kzr  \citep{Gillanders2020} or AT2018cow \citep{Metzger2022}, as well as long gamma-ray bursts (GRBs) such as GRB 211211A \citep{Yang2022,Zhong2023} or GRB 060614 \citep{King2007}. NS-WD mergers are also predicted to result in magnetars, and  were suggested to cause some fast radio bursts (FRBs) like FRB 190824 and FRB 180916.J0158+65 \citep{Zhong2020b}.

In this work, we aim to study such a merger by the means of magnetohydrodynamic (MHD) simulations.
Previous studies have looked into several different aspects of these mergers. \citet{Fernandez2019} and \citet{Zenati2019} used two-dimensional hydrodynamic simulations to study the accretion disk resulting from the disruption of the WD, and the nuclear reactions that would take place on it. \citet{Paschalidis2012} performed the first fully general relativistic simulations of mergers between NSs and scaled-down ``pseudo-WDs'' and studied whether the remnant would collapse into a BH. \citet{Zenati2020} and \citet{Bobrick2022} ran three-dimensional (3D) simulations of NS-WD mergers using smooth particle hydrodynamics (SPH) starting from realistic initial conditions.
The enormous range of scales required for this simulations, spanning over 10 orders of magnitude in density, forced previous works to revert to two-dimensional models or a  relatively low resolution when simulating these events. Furthermore,
none of the existing hydro simulations have yet included magnetic fields, which could be decisive in shaping the merger and post-merger dynamics \citep{Zhong2020b}. 
In this paper, we study the dynamics of the disruption and present the first high resolution, 3D MHD simulation of a NS-WD merger.

\section{Methods\label{sec:method}} 
For our simulations, we use the \AREPO code \citep{Springel2010, Pakmor2011, Weinberger2020, Pakmor2016}.
\AREPO uses the finite volume method to solve the 3D, fully compressible equations of ideal MHD with gravity,
\begin{equation}\label{continuity-eq}
    \frac{\partial \rho}{\partial t} + \nabla \cdot (\rho \mathbf{u}) =  0 ,
\end{equation}
\begin{equation}    
   \frac{\partial  (\rho \mathbf{u})}{\partial t} + \nabla \cdot  ( \rho \mathbf{u}
   \otimes \mathbf{u} + p_\mathrm{tot} \mathbb{I} - \mathbf{B} \otimes \mathbf{B}) =  \rho \mathbf{g}, 
\end{equation}
\begin{equation}
     \frac{\partial E}{\partial t} + \nabla \cdot [(E + p_\mathrm{tot} )
   \mathbf{u} - \mathbf{B}(\mathbf{B} \cdot \mathbf{u}) ]  =  \rho (\mathbf{g}\cdot\mathbf{u}), 
\end{equation}  
\begin{equation}\label{induction-eq}
   \frac{\partial  \mathbf{B}}{\partial t} + \nabla \cdot  ( \mathbf{u}
   \otimes \mathbf{B} - \mathbf{B}
   \otimes \mathbf{u} )  =  \mathbf{0},
\end{equation}
where $\rho$ denotes the density, $\mathbf{u}$ the velocity,  $\mathbf{B}$ the magnetic field\footnote{Here we use the Lorentz-Heaviside notation, $\mathbf{B}=\mathbf{b}/\sqrt{4\pi}$.},  $p_\mathrm{tot}=p+|\mathbf{B}|^2/2$ the total pressure, $p$ the gas pressure, $E$ the total energy density per unit volume, and $\mathbf{g}$ the gravitational acceleration. Equations~(\ref{continuity-eq})-(\ref{induction-eq}) are discretized on an unstructured, moving, Voronoi mesh as described in \citealt{Pakmor2016}.
The boundary conditions are periodic for all variables except the gravitational potential, which we compute by assuming that the density outside of the box domain is $0$.
Hyperbolic fluxes are computed at cell interfaces using the five-wave HLLD solver of \citet{Miyoshi2005}, and the strength of magnetic monopoles is kept under control using the Powell scheme \citep{Powell1999}. 
The grid is generated each timestep from a set of mesh-generating points that move with the local velocity of the fluid with small corrections to keep the mesh regular, leading to a nearly Lagrangian scheme. The resolution of the simulation can be changed locally by refining or de-refining the mesh following a set of criteria. The main criterion enforces the same mass for all cells within a factor of two of a chosen target mass and a maximum volume ratio between neighboring cells of a factor of five. 
On top of these criteria, we employ additional refinement criteria which are described in detail in Sec.~\ref{sec:setup} and Sec.~\ref{sec:binsetup}.
Self-gravity is included with a tree-based algorithm for the gas cells, and via exact force computation for the NS. We perform these simulations using Newtonian gravity, as the processes and dynamics that we aim to resolve are sufficiently distant from the NS. The equation of state (EoS) used in our simulations is the Helmholtz EoS \citep{Timmes2000}. We assume that the simulation takes place in a regime where the gas is completely transparent to neutrinos, and they leave the system without interactions. Therefore, neutrino transport simplifies to a cooling process. We take into account thermal neutrino cooling by implementing\footnote{Adapting Frank Timmes' scripts which can be found in:\\ \url{http://cococubed.asu.edu/code_pages/nuloss.shtml}} the fits from \citet{Itoh1996a}.
The energy lost to GW is negligible and not taken into account, as it is not dynamically relevant for the part of the merger that we simulate.  We compute the GWs in the same way as \citet{Moran-Fraile2023} by calculating the approximate quadrupole radiation from Newtonian gravity.
We do not use a nuclear network as nuclear processes are predicted to play a minor role in the dynamical evolution of the disruption \citep{Zenati2019}.

\subsection{Stellar setup\label{sec:setup}}
As a primary star in our binary system we take a 1.4\msol NS and for the companion we take a 1\msol WD composed of an equal-by-mass mixture of carbon and oxygen, with a central density of $\rho_\textrm{c} = 3.4\E{7}\textrm{g\,cm}^{-3}$ and an initial radius of $5.3\E{8}\,\textrm{cm}$ (defined as the radius containing 99.9\% of the stellar mass). The WD model is generated with a uniform initial temperature of $5\E{7}\,\textrm{K}$. We then obtain the density profile by assuming hydrostatic equilibrium, and map it to 3D employing a healpix based algorithm following \citet{Pakmor2012} and \citet{Ohlmann2017}. We use the Helmholtz EoS to compute the values of the internal energy for the \AREPO cells given the temperature. We add an initial dipole magnetic field with an polar surface value of $10^{5}\,\textrm{G}$ to the WD. The magnetic dipole field is  set up along the z-axis and we impose a cutoff radius in the center of the star with a size equal to two times the radius of the smallest \AREPO cell, to avoid an arbitrarily large value for the magnetic field in the center of the WD.

This WD is relaxed in isolation to ensure hydrostatic equilibrium, as deviations from equilibrium could appear due to discretization errors when mapping the one-dimensional profiles to the unstructured 3D mesh of \AREPO. For the relaxation we place the star in a box of $10^{10}\,\textrm{cm}$ side length with a uniform background grid of density $\rho=10^{-5}\,\textrm{g\,cm}^{-3}$. The total relaxation time is chosen to be ten times the dynamical timescale of the star,  corresponding to $11\,\mathrm{s}$. During the first half of this time we apply a damping force  to reduce the spurious velocities that could have arisen during the mapping. During the second half of the relaxation, the star is left to evolve on its own (with no damping force being applied), and we check that the density, pressure, and internal energy profiles remain stable.

Due to the difference in the dynamical timescales of a NS and a WD, it is computationally unachievable to resolve the hydrodynamic evolution of both the NS and the WD in one simulation.
We therefore model the NS as a gravitation-only point particle with a mass of $1.4$\msol.
To limit the pressure gradient next to our point mass NS, we use a softened gravitational potential.
We found that low resolution inside the softening length produces issues with energy conservation from the beginning of the simulation.
At the same time, increasing the resolution close to the NS becomes prohibitively expensive for small values of the softening length.
Experimentally we increased the resolution while decreasing the size of the softening length, and we found a good compromise between computational expense and energy conservation by using a softening length of $8.4\E{7}\,\textrm{cm}$ and enforce a minimum of 40 cells per dimension within the softening length.

The gravitational potential inside the volume enclosed by the softening length deviates from a Newtonian potential. 
Therefore, we consider this region unresolved and we report the values obtained by the simulation within this volume solely as a reference, as they may deviate from realistic ones.
As a point mass, the only characteristic for a NS that we assume is its mass and we do not model any other physical property. Thus we cannot give it a magnetic field with the current implementation of MHD in \AREPO.
We could set a dipole configuration in the background around it, but as the  mass contained in the background is so small compared with the one of the WD, as soon as accretion onto the NS starts, the initial magnetic field around the NS will become dynamically irrelevant .
Therefore, we did not establish an initial magnetic field around the NS. 

\subsection{Binary setup}\label{sec:binsetup}
The target mass resolution of our simulation is $m_\mathrm{cell} = 5\E{-7} \mathrm{M}_\odot$, which results in ${\sim}\,2\E{6}$ \AREPO cells before further refinement.
Due to the extreme pressure and density gradients in the neighbourhood of the NS, we further increase the resolution in a disk-like region around the NS where a large fraction of the gas will end up after the disruption of the WD. We define this region as a cylinder centred on the NS with a radius of $6\E{8}\,\textrm{cm}$ ($\sim$ half of the tidal disruption radius of the WD) and a height of $5\E{7}\,\textrm{cm}$. In this region, we impose the same refinement and derefinement conditions as within the softening length of the NS, which in turn means ensuring a minimum of $\sim 6$ million particles in this region. We allow for further refinement if cells inside this region get over our target $m_\mathrm{cell}$. With this setup, the typical total number of cells in the simulation is $\sim8.1$ million.

At the beginning of the simulation, the stars are placed in a Keplerian circular orbit at a distance of $2.27\E{9}\textrm{cm}$, which corresponds  to 1.5 times the separation at Roche-lobe overflow \citep[RLOF, ][]{Eggleton1983a}. The WD is initially set to solid-body rotation in co-rotation with the NS at the point of RLOF. The orbit is then artificially shrunk on a timescale larger than the dynamical timescale of the WD so that potential tidal effects can still take place. This is done using the scheme described by \citet{Pakmor2021} that removes angular momentum from the binary system.
We stop shrinking the orbit at the point of RLOF (corresponding to an orbital separation of $1.52\E{9}\textrm{cm}$), when a steady mass transfer sets in.
At this point, the system has a total angular momentum of $L=8.3\E{50}\textrm{g\,cm}^2\,\textrm{s}^{-1}$.
From here on the system is left to evolve on its own and the WD disruption takes place on a few dynamical timescales. 
As the outer layers of the WD cannot be fully resolved, there is some numerical mass loss during the orbit-shrinking stage.
Part of this mass will be accreted by the NS before the onset of the steady accretion stream.
Furthermore, as explained in \citet{Bobrick2017}, the tidal force exerted by the NS deforms the WD making it not completely spherical and its gravitational potential not perfectly Keplerian. This results in our initial orbit having a small eccentricity ($\sim 0.001$), which also contributes to a small amount of mass transfer before the end of the orbit shrinking process.
The amount of mass accreted into the  softening length at the time that we stop the orbit shrinking is $10^{-4}\,\textrm{M}_\odot$. This initial mass transfer is dynamically irrelevant but during this stage the amplification of the magnetic fields around the NS already begins.

Table \ref{tab:ic} collects the initial parameters and important timescales of the numerical setup for easy reference.

\section{Results} \label{sec:results}
\begin{table*}[]
\caption{
Mass of the NS ($M_\textrm{NS}$), mass of the WD ($M_\textrm{WD}$), orbital separation after the end of the artificial orbital shrinkage ($a_0$), orbital separation at which the WD is tidally disrupted ($a_\textrm{disruption}$), time elapsed until the disruption of the WD ($t_\textrm{disruption}$), jet launching ($t_\textrm{jet, launch}$) and end of the simulation ($t_\textrm{fin}$); total magnetic energy ($E_{B,0}$) and average magnetic field strength ($\langle B_\textrm{0}\rangle$) at the beginning of the simulation and the value of the dipole magnetic field seed at the equator of the WD ($B_\textrm{seed}$).}
    \label{tab:ic}
    \centering
    \begin{tabular}{c|c|c|c|c|c|c|c|c|c}\toprule
        $M_\textrm{NS}$ & $M_\textrm{WD}$  & $a_0$ & $a_\textrm{disruption}$ & 
        $t_\textrm{disruption}$ & $t_\textrm{jet, launch}$ & $t_\textrm{fin}$ & 

        $E_{B,0}$ & 

        $\langle B_\textrm{0}\rangle$ & $B_\textrm{seed}$ \\
        
        [\msol] & [\msol] & [$\textrm{cm}$] & [$\textrm{cm}$] &
        [$\textrm{s}$] & [$\textrm{s}$] & [$\textrm{s}$] &
        [$\textrm{erg}$] & [$\textrm{G}$] & [$\textrm{G}$] \\
        \midrule

        $1.4$ & $1$ & $1.52\E{9}$ & $1.14\E{9}$ &
        $\sim80$ & $\sim140$ & $242$ & $2\E{41}$ &
          $2.5\E{7}$ & $10^{5}$ \\
         \bottomrule
    \end{tabular}
    
\end{table*}

\begin{figure*}
    \includegraphics[width=\textwidth]{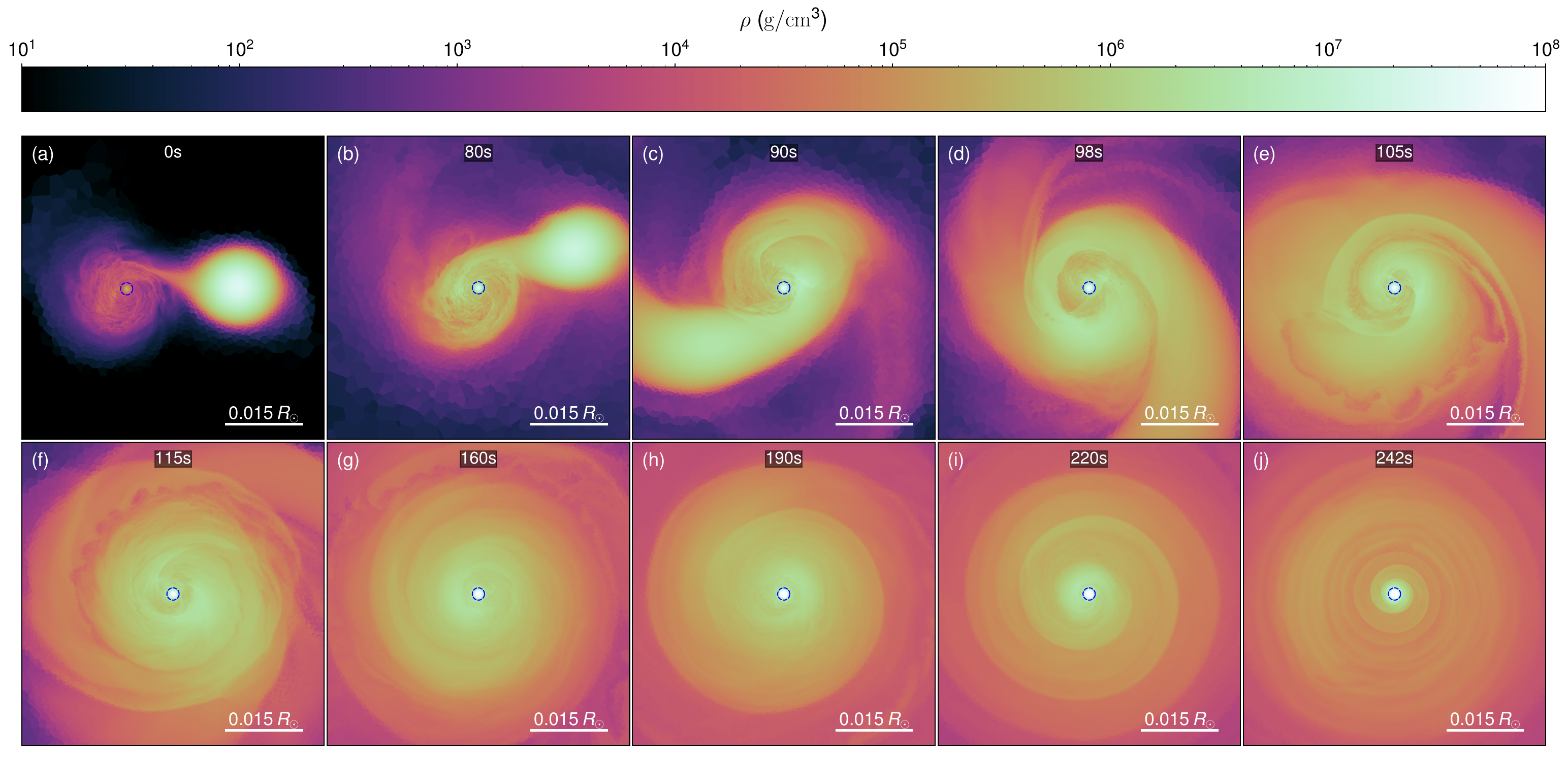}
    \caption{\label{fig:evolution}Hydrodynamical evolution of the merger. The panels show density slices in the orbital plane from the beginning of the simulation (panel a) until we stop it at $t=242\,\textrm{s}$ (panel j). The region with a softened gravitational potential is shown in a blue dashed line. }
\end{figure*}

\begin{figure}
    \centering
    \includegraphics[width=\columnwidth]{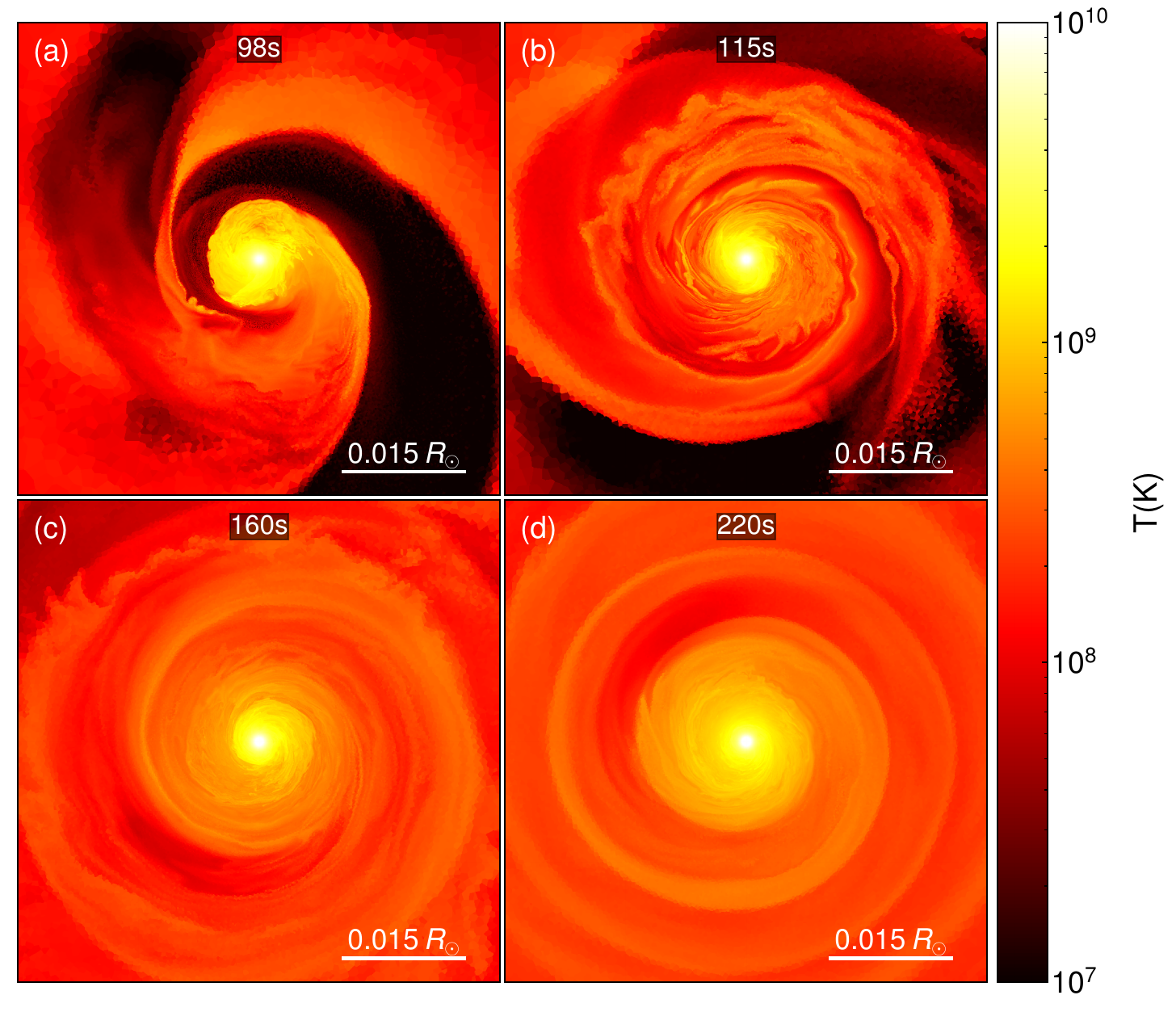}
    \caption{Slice through the orbital plane showing the temperature at selected times.}
    \label{fig:temp}
\end{figure}

\begin{figure}
    \centering
    \includegraphics[width=\columnwidth]{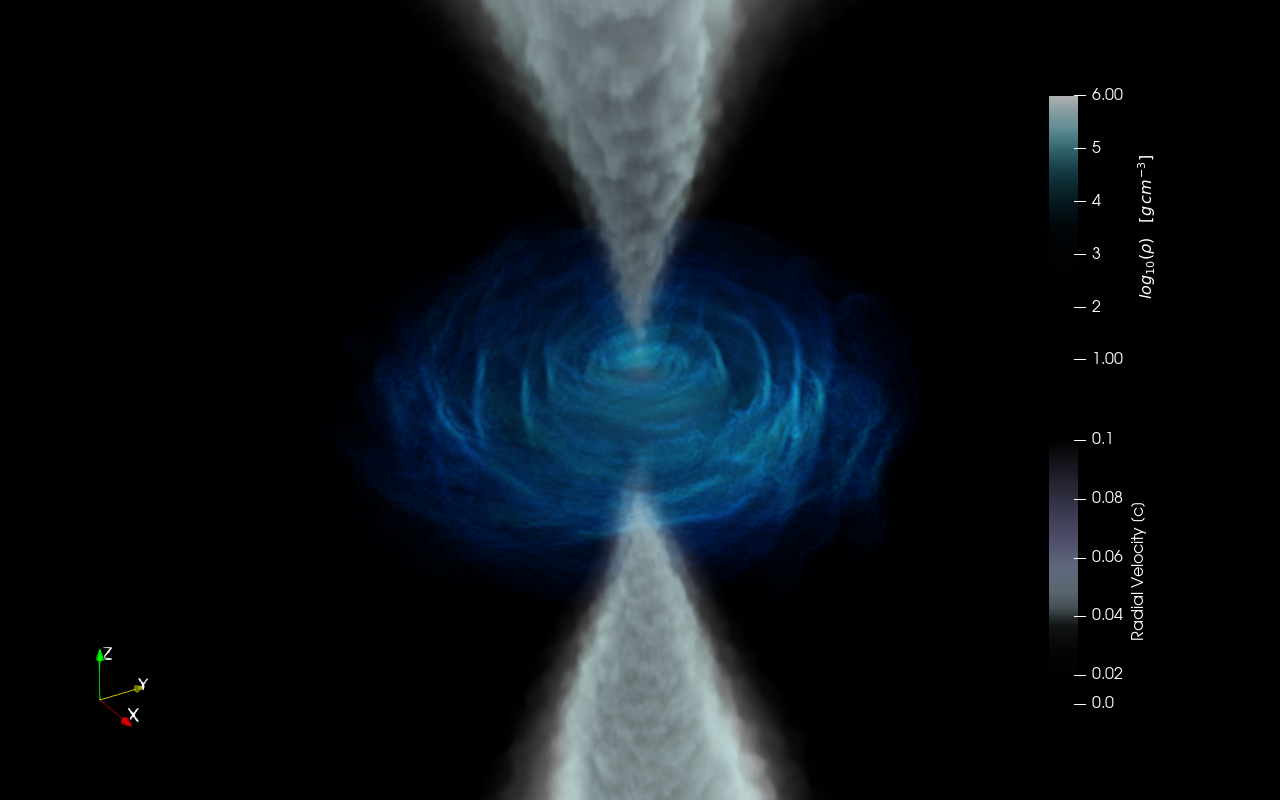}
    \caption{Movie (\href{https://jmoran2611.github.io/publications/nswd2023/}{online}) showing a 3D volume rendering of the density and radial velocity in the system. For visualization purposes, densities below $\rho < 10^3\,\textrm{g\,cm}^{-3}$ and radial velocities below $v_r < 0.03\textrm{c}$ are completely transparent. }
    \label{fig:3drender}
\end{figure}

The hydrodynamical evolution of the merger is summarised in Fig.~\ref{fig:evolution}\footnote{Additional movies are available in \url{https://doi.org/10.5281/zenodo.8073873}} and the temperature is shown in Fig.~\ref{fig:temp}. A movie showing a 3D rendering of the merger is also available in Fig.~\ref{fig:3drender}.
After the end of the artificial orbital shrinkage ($t=0\,\textrm{s}$), a steady mass transfer has developed. Moreover, during this stage, a small fraction of mass from the WD leaves the binary through the Lagrange point $L_2$.
The transfer of mass and angular momentum makes the orbit shrink further, and after four complete orbits ($t=80\,\textrm{s}$, Fig.~\ref{fig:evolution}b), the WD gets closer to the NS than its tidal disruption radius and is torn apart at an orbital separation of $1.14\E{9}\,\textrm{cm}$. By this point, $0.005$\msol of gas has been unbound.
This takes place mainly through the $L_2$ point, with a smaller fraction of the gas being also ejected along the polar axis of the NS. 
The disruption unbinds another $0.01$\msol of material.
The disrupted stellar material initially forms a spiral arm and then starts to be accreted onto the NS (Fig.~\ref{fig:evolution}d).

The infalling material intersects itself while accreting, developing shear instabilities and shocks that heat the gas, and transport angular momentum outwards. In this transient process, the orbit of the disrupted material circularises and a disk is formed due to the tidal interaction of the turbulent, infalling material with the NS (Fig.~\ref{fig:evolution} e and f). The temperature in the shocks  reaches values up to $T=9\times10^8\,\textrm{K}$ (Fig.~\ref{fig:temp}), but the hottest regions of the shocks remain at rather low densities of   $<10^6\,\textrm{g\,cm}^{-3}$. Therefore, only minor nuclear burning is expected to take place in these shocks. The regions close to the NS heat up rapidly, reaching temperatures of $10^{10}\,\textrm{K}$ inside the softening length. In this region nuclear reactions are essentially unavoidable. By $t\simeq160\,\textrm{s}$, the disrupted material has formed a disk-like structure around the NS.
As the NS does not have a surface and its gravitational potential is softened, the accreted mass accumulates inside the volume of the softening length. In our simulation, any ``acretion'' is not onto the NS, but  into this volume.
During the $150 \, \mathrm{s}$ following the disruption of the WD ($t < 230\,\textrm{s}$), $0.056$\msol of material is accreted into the gravitationally softened volume at an average rate of $3.7\E{-4}\,\textrm{M}_\odot\,\textrm{s}^{-1}$. The density inside this region keeps increasing as material is accreted. At the time of the WD disruption, the maximum density is $\rho =2\E{7}\,\textrm{g\,cm}^{-3}$, and at the end of the simulation it has reached a value of $\rho= 1.4\times10^9\,\textrm{g\,cm}^{-3}$.
As we do not model nor resolve the physical processes around and at the NS, we observe an artificial accumulation of material inside the softened volume. Part of this material should be accreted onto the NS, potentially through an accretion disk.
During several stages of the simulation the combination of temperatures and densities inside the softened volume are suitable for a strong nuclear burning, which would also impact the density and temperature profiles in the neighbourhood of the softening length.

\begin{figure}
    \includegraphics[width=\columnwidth]{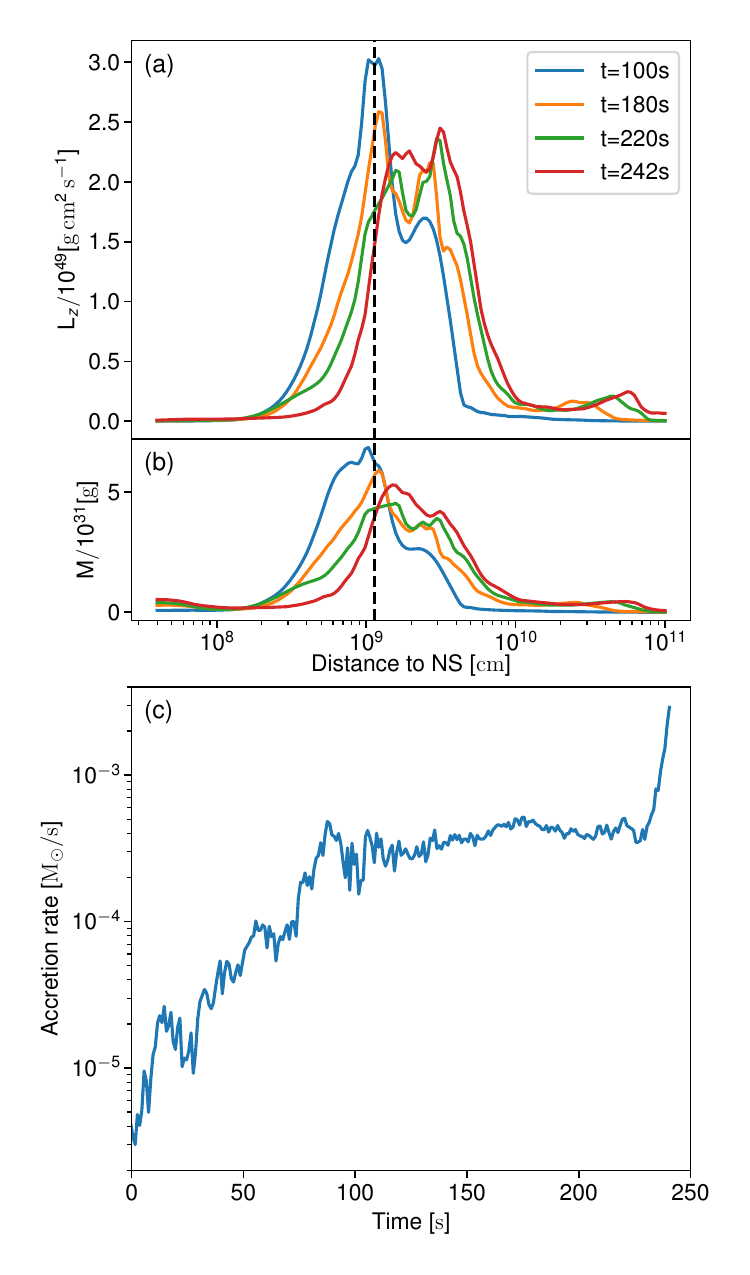}
    \caption{\label{fig:Lz}Panel a: Z-component of the angular momentum with respect to the distance to the NS for selected times. Panel b: Distribution of gas mass with respect to the distance to the NS for selected times. The distance at which the WD is disrupted is shown by the black dashed line. 
    Panel c: Rate in \msol$\textrm{s}^{-1}$ at which mass is accreted onto the gravitationally softened volume.}
\end{figure}

The tidal interaction of the NS with the non-uniform accretion disk generates shocks, producing spiral density waves (Fig.~\ref{fig:evolution}i and Fig.~\ref{fig:evolution}j) that transfer angular momentum (Fig.~\ref{fig:Lz}a). The heat dissipation produced by the shocks speeds up the transfer of angular momentum, increasing the accretion rate (Fig.~\ref{fig:Lz}c). This culminates in a rapid accretion event at $t=242\,\textrm{s}$, where we observe a peak accretion rate of $3\E{-3}\,\textrm{M}_\odot\,\textrm{s}^{-1}$. At this point, $0.077$\msol of gas has been accreted into the gravitationally softened volume. 
Inside the gravitationally softened volume, the gas rotates as a solid body and transitions to a quasi-Keplerian rotation in the accretion disk.
The distribution of angular momentum and mass during the simulation can be seen in Fig.~\ref{fig:Lz}a-b. Most of the mass and angular momentum after the disruption of the WD is concentrated at a distance from the NS equal to the tidal disruption radius of the WD. It is initially slowly pushed out of this region by the turbulent motions and the shocks. When the spiral waves are developed they transport momentum outwards very efficiently, as can be seen by the rapid change in the angular momentum distribution between $t=180\,\textrm{s}$ and $t=242\,\textrm{s}$.

At $t=242\,\textrm{s}$ we stop the simulation as we cannot resolve the accretion and energy errors start accumulating in the simulation. Considering every \AREPO cell with a total energy larger than zero as unbound material, by the time we end the simulation, $0.94$\msol of gas remain gravitationally bounded when only taking into account potential and kinetic energy. This number decreases to $0.93$\msol when also including magnetic and internal energy in the total energy.

\subsection{Magnetic field amplification}
\begin{figure*}
    \centering
    \includegraphics[width=\textwidth]{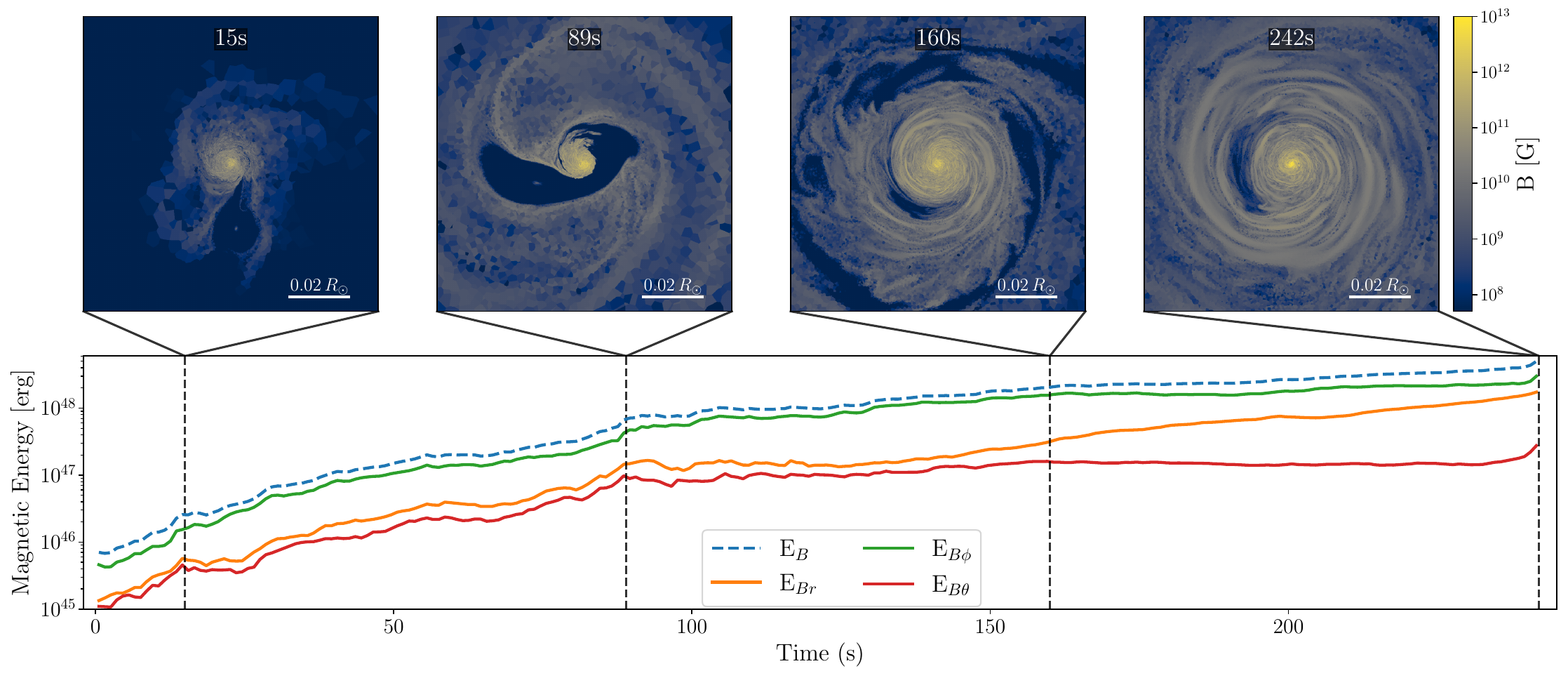}
    \caption{\label{fig:bfld}Evolution of the magnetic field throughout the simulation. \emph{Top}: Slices through the orbital plane, centered on the NS, showing in color the absolute magnitude of the magnetic field for selected snapshots.
    \emph{Bottom}: Evolution of the total magnetic energy over time (blue, dashed line) as well as the energy of the different components of the magnetic field in spherical coordinates (orange, green and red for the radial, azimuthal and polar components, respectively).}    
\end{figure*}

The evolution of the magnetic energy in the simulation is shown in Figure~\ref{fig:bfld}.
As explained in Sec.~\ref{sec:setup}, the WD is given an initial magnetic field seed with a value of $10^{5}\,\textrm{G}$ at the equator. As soon as mass transfer begins, shear within the accretion stream amplifies the magnetic field around the NS, while compression of the magnetic field lines amplifies it in the shocked regions.
The amplification process is complex and involves a combination of several mechanisms including magneto-rotational instability (MRI, discussed in more detail in Appendix \ref{sec:appendix1}), Kelvin-Helmholtz instabilities and dynamo processes, similar to what is observed in 3D simulations of common-envelope evolution \citep{Ondratschek2022}, main-sequence star mergers \citep{Schneider2019}, binary white dwarf mergers \citep{Zhu2015} or binary neutron star mergers \citep{Obergaulinger2010, Kiuchi2015, Aguilera-Miret2020}.
This amplification process begins as soon as we start the simulation and the NS begins to accrete the pseudo-vacuum surrounding it as well as part of the mass lost by the WD during this stage. By the time the artificial orbital shrinking phase finishes, the magnetic energy (\eb) of the system has increased from an initial value of $2\E{41}\,\textrm{erg}$ up to a value of $7.0\E{45}\,\textrm{erg}$, with the fields reaching values of up to $5.6\E{11}\,\textrm{G}$ inside the gravitationally softened volume around the NS.

This initial amplification of the magnetic energy due to the NS accreting the pseudovacuum is unavoidable. To ensure that the amplification is not boosted by our artificial orbit shrinkage, we have performed two more simulations.
In both of them we have placed the stars at a distance of $1.1$ times the distance of RLOF.
In the first one, we let the stars evolve freely (keeping constant orbital separation), and in the second one we shrink the orbit further down to the distance of RLOF before letting it evolve freely.
At this initial distance of $1.1$ times the separation at RLOF, the accretion of mass lost from the WD by numerical effects is the main source of magnetic amplification rather than the accretion of the pseudo-vacuum.
In both simulations, the total magnetic energy is amplified to values of $\sim1\E{46}\,\textrm{erg}$ on a timescale of $25\,\textrm{s}$, matching the magnetic energy generated in our fiducial simulation at this separation but generating it in a shorter timescale. The simulation left to evolve freely at a separation equal to $1.1$ the separation at RLOF remains at a constant orbital separation, and after this initial fast amplification, the growth of the magnetic energy slows down and plateaus around $2\E{46}\,\textrm{erg}$.
In contrast, in the simulation where we shrink the orbit down to the point of RLOF, a steady mass transfer sets in and the magnetic fields continue to be amplified with the evolution of the total magnetic energy matching the one of our fiducial simulation.
Hence we conclude that this initial phase of magnetic amplification previous to the start of our simulation is self-consistent and independent of the artificial orbital shrinking.

The amplification of the magnetic field takes place dominantly in a toroidal direction around the NS, following the accretion stream, as evidenced by the $\phi$ component of the magnetic energy in Fig.~\ref{fig:bfld}.
When a steady mass transfer sets in, the amplification of the magnetic field speeds up as the NS begins to accrete the outermost layers of the WD. At this initial stage, the \emph{e}-folding growth time of the magnetic energy peaks at a value of $\sim20\,\textrm{s}$.
As the orbit shrinks and the mass transfer rate increases, the magnetic energy keeps steadily growing, with an abrupt increase shortly after the disruption of the WD due to the turbulent motions.
At this point, the total magnetic energy has increased to $6.7\E{47}\,\textrm{erg}$, reaching maximum local values of up to $5\E{12}\,\textrm{G}$ inside the gravitationally softened region around the NS.
During the period of rapid accretion onto the NS driven by the spiral waves ($t\sim240\,\textrm{s}$), there is a second phase of fast amplification of the magnetic fields with \eb raising up to $5\E{48}\,\textrm{erg}$ and reaching maximum values of $2.3\E{13}\,\textrm{G}$ next to the NS.

\begin{figure}
    \centering
    \includegraphics[width=\columnwidth]{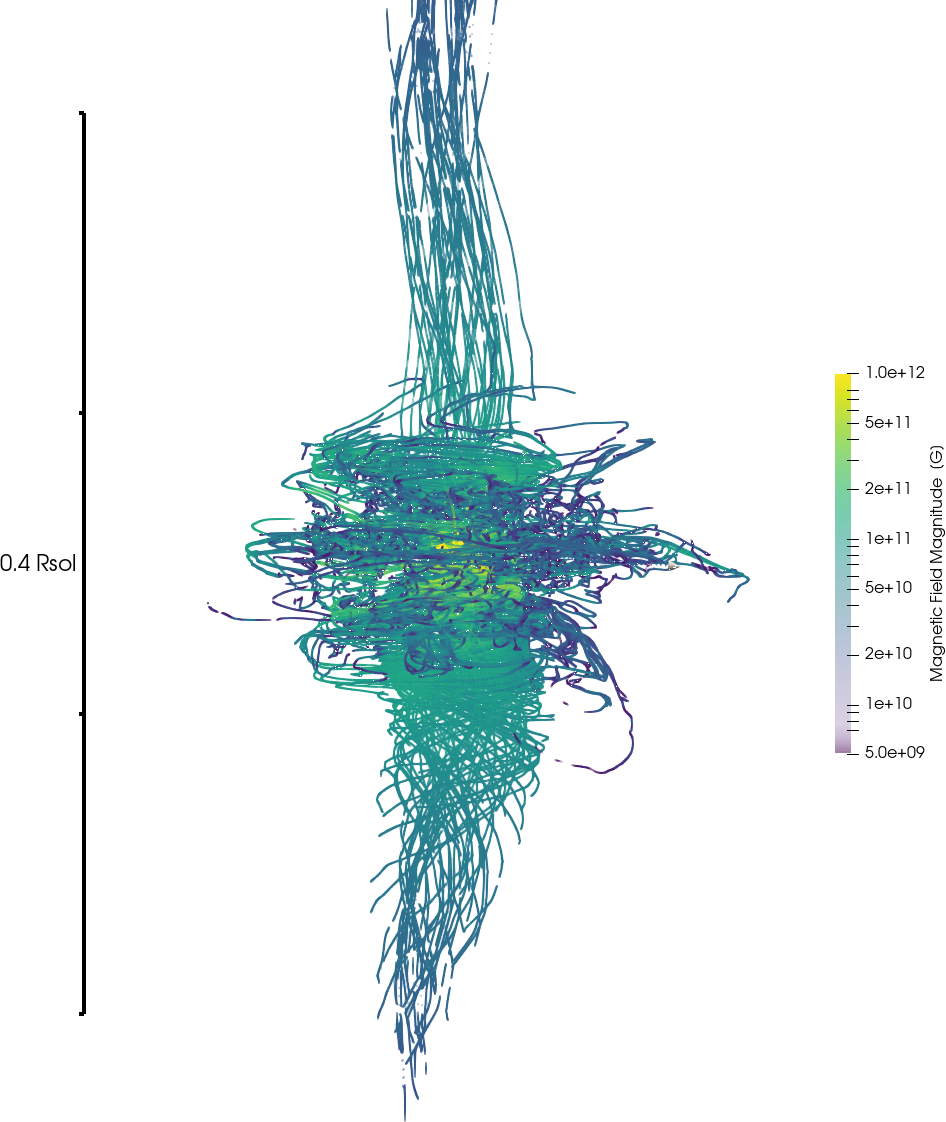}
    \caption{3D rendering of magnetic field lines at $t=200\textrm{s}$}
    \label{fig:bfield_lines}
\end{figure}

\subsection{Jet-launching}
\begin{figure*}
    \includegraphics[width=0.97\textwidth]{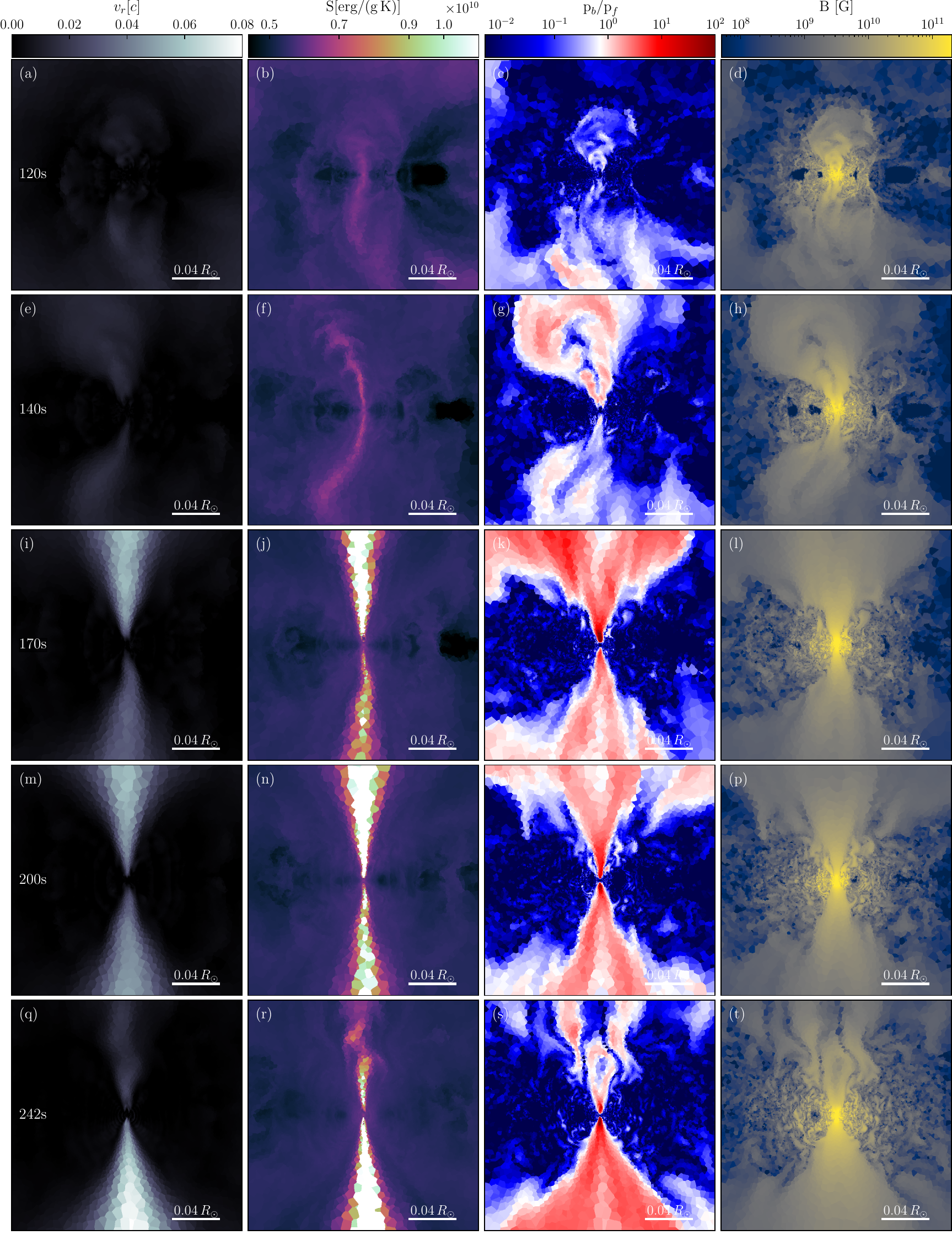}
    \caption{\label{fig:jet}Rows show slices through the midplane perpendicular to the orbital plane of the binary, centered on the NS before and after the formation of the polar outflows. The first column shows the radial velocity of the gas as a fraction of the speed of light, ($v_r/c$). The second column shows the entropy of the gas. The third column shows the magnetic-to-fluid pressure ratio, $p_b/p_f$. The fourth column shows the absolute magnitude of the magnetic field in the gas.}
\end{figure*}
\begin{figure}
    \centering
    \includegraphics[width=\columnwidth]{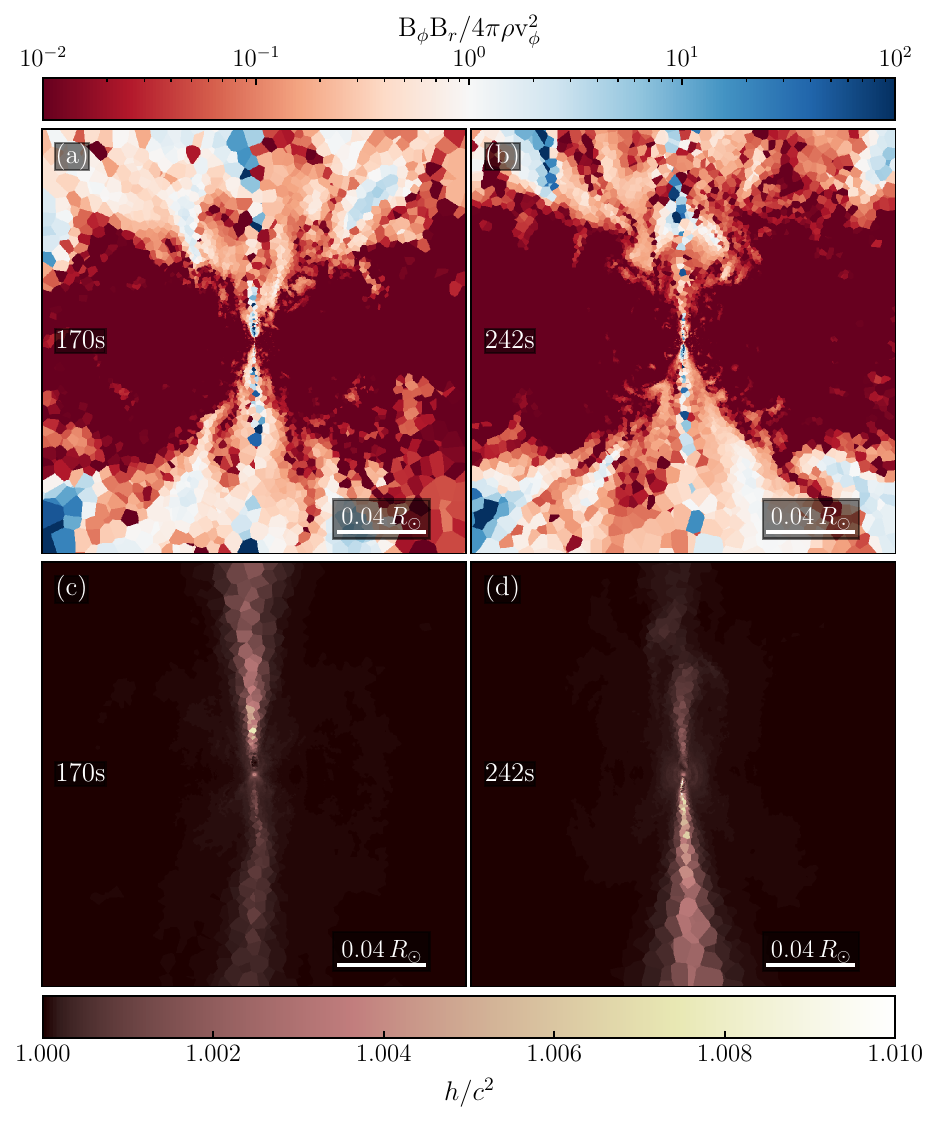}
    \caption{Slices through the midplane perpendicular to the orbital plane of the binary, centered on the NS at two times during the jet evolution. The top row shows the  $r \phi$ component of the Maxwell stress tensor in the gas normalized by the centrifugal component of the Reynolds stress tensor. The bottom row shows the specific enthalpy of the gas}
    \label{fig:jet_extra}
\end{figure}
\begin{figure}
    \centering
    \includegraphics[width=\columnwidth]{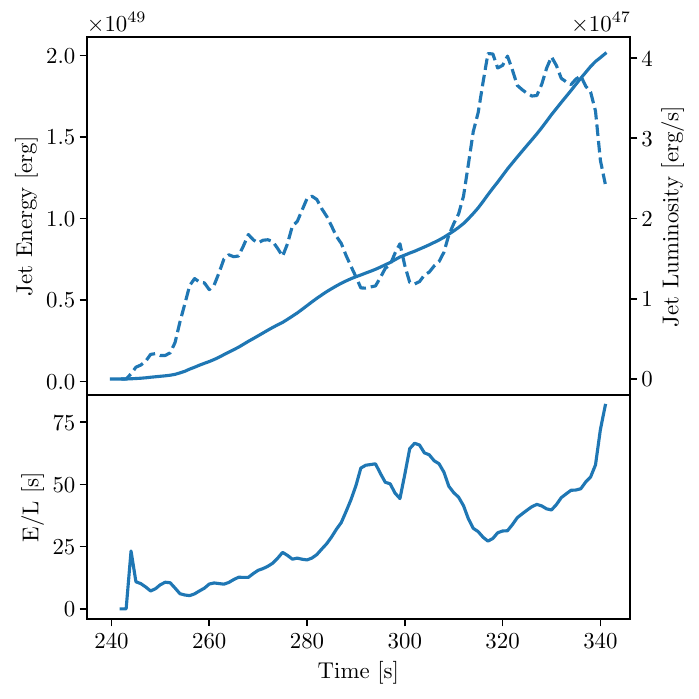}
    \caption{Time evolution of energy and luminosity of the jets. Top: energy in the jets is shown by the plain blue line and the jets luminosity is shown in a blue dashed line. Bottom: energy divided by luminosity over time.}
    \label{fig:jet_luminosity}
\end{figure}
The magnetic fields generated throughout the entire event power two polar outflows launched in opposite directions (Fig.~\ref{fig:jet}). A 3D rendering of the magnetic field lines showing the toroidal structure that gives rise to the bi-polar outflows can be seen in Fig.~\ref{fig:bfield_lines} .
After the onset of mass transfer, regions with a magnetic-to-fluid pressure ratio ($p_b/p_f$) larger than one develop perpendicular to the orbital plane (third column in Fig.~\ref{fig:jet}), where the magnetic pressure exceeds the dynamic pressure. In these regions, the Maxwell stresses dominate the dynamics, driving the collimated ejecta (note the white-to-blue shades in the upper panels of Fig.~\ref{fig:jet_extra}).
As a result, since the beginning of the simulation, a small fraction of the accreting mass is ejected along the polar axis of the NS following the direction of the steepest magnetic pressure gradient.
This outflow becomes faster and more collimated as the energy associated with the radial component of the magnetic field ($\textrm{E}_{Br}$ in Fig.~\ref{fig:bfld}) increases. 
At $t\,{\sim}\,140\,\textrm{s}$ (Fig.~\ref{fig:jet}e--h) the outflows become distinguishable as highly magnetized, high-entropy gas  ejected perpendicular to the orbital plane with velocities close to $1\%$ of the speed of light $c$.
By $t\,{\sim}\,170\,\textrm{s}$ (Fig.~\ref{fig:jet}i--l), a bi-polar jet structure has developed with characteristic velocities of ${\sim}\,0.05c$, reaching maximum velocities of $0.08c$. In spite of the large entropy of the jets, the (relativistic) specific enthalpy, $h=1+\epsilon/c^2 + p_f/(\rho c^2)$ (where $\epsilon$ is the internal energy per unit mass), is very close to unity, indicating the ejection of a cold and highly magnetized jet.
The computed values of $h$ provide a proxy for the asymptotic Lorentz factor of the fluid, $\Gamma_\infty\,{\sim}\,1.005 - 1.01$ ($\Gamma_\infty\,{\sim}\,h$, according to Bernoulli's law along streamlines). Likely, this is a lower bound, since after the formation of a low-density polar funnel, matter further ejected through the jet will be accelerated to higher terminal speeds. The strength of the jets varies in time and evolves unevenly in the two directions.
During the late period of accretion dominated by the spiral waves, the outflow above the orbital plane weakens while the outflow below the orbital plane greatly increases in strength with typical velocities $v\sim0.08c$ and a maximum velocity of $v\sim0.1c$ at $t=242\,\textrm{s}$ (Fig.~\ref{fig:jet}q--t).
The outflows are initially launched at a distance of ${\sim}\,4.4\E{8}\,\textrm{cm}$ from the NS, which roughly corresponds to five times its softening length.
In the ejecta, $p_b/p_f\,{\sim}\,10$, and locally $p_b/p_f$ reaching values of ${\sim}\,1000$ close to the gravitationally softened volume around the NS.
The energy and luminosity of the jets is shown in  the upper panel of Fig.~\ref{fig:jet_luminosity}. There we measure a non-monotonically growing jet energy along with a growing luminosity, with imprinted variability timescales down to ${\sim}\, 25\,$s (Fig.~\ref{fig:jet_luminosity} bottom panel, showing the ratio jet-energy-to-jet-luminosity, which roughly informs about the timescale of changes in the jet luminosity).
Likely, this timescale is an upper bound of the smallest variability timescales induced by the jet interacting with the surrounding, stratified medium, dictated by the time it takes Alfv{\'e}n or sound waves to cross the jet transverse radius (${\sim}\,1$--$2\,$s; see, e.g., \citealt{Aloy2002a, Matsumoto2017} for analogous jets propagating in stellar environments).
Capturing the jet--environment interaction requires a very large numerical resolution on the jet, which we miss here, as outside of our disk-like region surrounding the NS the resolution criteria falls back to target a mass per cell of $m_\mathrm{cell} = 5\E{-7} \mathrm{M}_\odot$. Also, in our case, it depends on the operative definition of matter belonging to the jet. While the impact of small variations of the radial velocity threshold adopted to define the jet on its energy is negligible, the choice of the threshold may induce moderate changes for the luminosity variability timescale. Shorter timescales may be induced by the dynamics of the accretion/ejection of matter in the vicinity of the (here unresolved) NS, but likely at later times (in analogy to what is found for jets induced in collapsars \citealt{Aloy2000,  Matzner2003, Morsony2010, Gottlieb2020}). The jets eject material at an average rate of ${\sim}\,3\times10^{-4}\,\textrm{M}_\odot\,\textrm{s}^{-1}$, and have opening half-angles oscillating between ${\sim}\,22^\circ$ and $25^\circ$ for the top jet and between $18^\circ$ and $22^\circ$ for the bottom jet.

As the accretion rate increases, the region of gas with the highest $p_b/p_f$ draws closer to the gravitationally softened volume around the NS,
and, by $t\,{\sim}\,242\,\textrm{s}$, the material is launched from the outer regions of the softened volume.

As gravity inside the softening length around the NS diverges from a realistic gravitational potential, we must be careful when interpreting processes that have an origin close to the softened volume. 
Also, nuclear reactions can massively change the energy budget of the central region.
As the jet is initially launched at a reasonably large distance from the NS, we consider its formation a realistic physical effect rather than an artifact of the numerical model.
However, as the region of acceleration draws closer to the NS, we must consider the jet-launching at later stages unresolved.
Moreover, we use a non-relativistic version of \AREPO and therefore gas moving at velocities close to $0.1c$ is no longer accurately treated, adding to our conclusion to stop the simulation during the fast accretion event that takes place at $t\,{=}\,242\,\textrm{s}$.

\subsection{Gravitational Waves}
\begin{figure}
    \centering
    \includegraphics{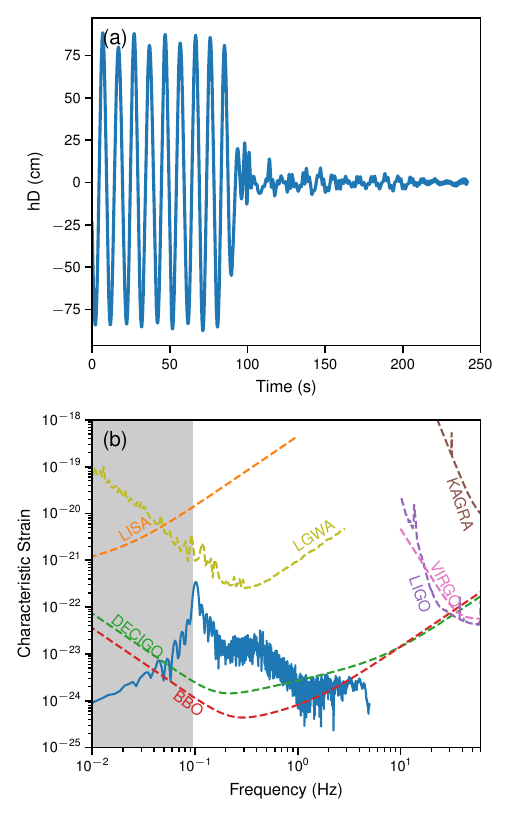}
    \caption{Gravitational waves produced in the simulation. Top: strain of the GW signal $h$ times the distance D to the source, observed from the z-direction (polar axis), over time.
    Bottom: Characteristic strain of the GW signal assuming a distance of 1Mpc to the source along the polar axis. The sensitivities of LISA \citep{Robson2019}, DECIGO and BBO \citep{Yagi2011}, LIGO, VIRGO and KAGRA \citep{LVK2023} and LGWA\citep{Dupletsa2023} are shown as dashed orange, green, red, purple, brown and yellow lines respectively. The gray shaded area denotes the frequencies below two times our initial orbital frequency where signal coming from the inspiral phase of the binary is expected to be present.
    }
    \label{fig:gws}
\end{figure}

Figure~\ref{fig:gws} shows the GWs released during the simulation as seen from the top of the orbital plane, which is the viewing angle that results in the loudest signal, providing a best-case scenario for their observation. The cross ($h_\times$) and plus ($h_+$) polarizations of the gravitational waves have been computed using the approximate quadrupole radiation in the same way as in \citet{Moran-Fraile2023}, and the characteristic strain is defined as $h_c(f) = 2f\Tilde{h}(f)$ where $\Tilde{h}$ is the Fourier transform of the time-domain amplitude of the GW,
\begin{equation}
    \Tilde{h}(f) = \frac{1}{R}\sqrt{|\Tilde{h}_+(f)|^2 + |\Tilde{h}_\times(f)|^2},
\end{equation}
with 
\begin{equation}
    \Tilde{h}(f) = \int^\infty_{-\infty}e^{-2\pi i f t}h(t)\dd t .
\end{equation}
During the accretion of the outer parts of the WD, the amplitude of the waves oscillates in the time domain (Fig.~\ref{fig:gws}a) mainly due to the eccentricity of the binary orbit at this stage, but it does not significantly rise or decrease prior to the merger, while the frequency remains approximately constant. The emission reduces abruptly shortly after the tidal disruption of the WD. Only higher frequency components due to the lack of perfect axial symmetry of the accretion disk remain in the signal (Fig.~\ref{fig:gws}a). 

The characteristic strain (Fig.~\ref{fig:gws}b) shows a strong peak at $f=0.1\,\textrm{Hz}$, which roughly corresponds to twice the orbital frequency of the binary at $t=0\,\textrm{s}$ ($f_{\textrm{orb}}=4.8\E{-2}\,\textrm{Hz}$). As expected from binary star dynamics, the frequency of the GWs released corresponds to twice the orbital frequency of the system.
This can be used to explain some of the main features seen in the characteristic strain. Because we only simulate the merger from an initial orbital frequency of $\sim0.05\,\textrm{Hz}$, we miss the GWs released during the previous stages of the evolution of the binary, which would emit GWs mainly at frequencies below the peak. 
NS-WD binaries are expected to release GWs below this frequency during their inspiral stage \citep{VanZeist2023}.

In our simulation, the orbital frequency of the binary does not change substantially before the tidal disruption of the WD and the amplitude of the GWs decreases sharply after the disruption, explaining the narrow peak that we observe in the characteristic strain (Fig.~\ref{fig:gws}b). The higher-frequency components of the GW signal produced primarily after the disruption have a much lower amplitude than the GWs released prior to the merger. Between $f=0.2\,\textrm{Hz}$ and $f=0.5\,\textrm{Hz}$ they show a mainly constant characteristic strain $\sim50$ times smaller than at the peak. For higher frequencies, the strain drops down another order of magnitude, and at frequencies higher than $5\,\textrm{Hz}$ we become limited by the sampling rate of our simulation.
Frequencies on the order of Hz and higher may arise from material close to, and inside the softened volume, and should therefore be considered unresolved.

The characteristic frequencies of the signal released during and after the merger make this event an ideal target for deci-hertz observatories like BBO and DECIGO, which for a source at a distance of $1\textrm{Mpc}$ would produce a signal to noise ratio ($S/N$) of approximately 115 and 50, respectively. The ratio $S/N$ is defined as $$S/N = \sqrt{4\int^\infty_0\frac{\abs{\Tilde{h}(f)}^2}{S_n(f)}\dd f },$$ where $S_\mathrm{n}$ is the noise power spectrum density of the corresponding detector given by \citet{Yagi2011}. For the moon-based LGWA, the signal is not loud enough at $1\textrm{Mpc}$, and the merger would have to take place significantly closer to be detected.

\section{Discussion \& summary}
We have conducted the first 3D-MHD simulations of the merger between a $1$\msol COWD and a $1.4$\msol NS. We find that the disruption of the WD takes place after a few dynamical timescales since reaching the separation of RLOF. The mass transfer from the WD to the NS and the ensuing disruption of the WD amplify any pre-existing magnetic fields by several orders of magnitude, producing magnetic fields up to $10^{13}\,\textrm{G}$.

The strong magnetic fields generated during the event launch two mildly relativistic jets perpendicular to the orbital plane with velocities on the order of $0.05-0.1c$.
These values should be taken as order-of-magnitude estimate because several aspects of the final outcome of these mergers are expected to depend on the physics that are not resolved in our simulations. 

As we model the NS as a point mass particle with a softened potential, not only gravity inside the softened volume deviates from a realistic gravitational potential, but we also do not provide any physical model for accretion onto the NS, which would potentially be limited by the Eddington accretion luminosity, dramatically changing the conditions in the neighborhood of the NS.
As the conditions around the NS allow for nuclear burning, we have run a simulation in which we couple the MHD solver to a 13 isotope nuclear network following the same methods as in \citet{Gronow2021} and \citet{Pakmor2021}.
In this simulation, the production of $^{56}$Ni is observed both outside and inside the softening length.
While the first effect is resolved and physical, the burning inside the softening length is unresolved but may still be physical.
While we must exercise caution when interpreting these results, which will be studied in more detail in follow up publications, they do provide support for the idea that mergers between NS and WD result in iron rich transients such as AT2018kzr \citep{Gillanders2020}.

As reported in Sec.~\ref{fig:jet} the jet is initially launched sufficiently distant from the NS, so we consider its origin physical, but as accretion into the softened potential increases, the jet becomes faster and its acceleration region draws closer to the NS.
A smaller softening length will lead to larger flow accelerations close to the NS, potentially generating stronger magnetic fields, which may alter the properties of the jet. 
Moreover, because the velocities of the ejected material reach up to $0.1c$, our Newtonian treatment becomes less accurate and a relativistic solver is needed at some point \citep[e.g.][]{Lioutas2022}.

The merger of our two compact stellar remnants releases GWs in the deci-Hertz frequency range. We observe that GWs at the moment of WD disruption are released mainly at a frequency of $0.1\,\mathrm{Hz}$, and the post merger releases higher frequency GWs, making these mergers an ideal target for proposed  deci-Hertz observatories such as DECIGO, BBO and LGWA.
We report GWs up to frequencies of ${\sim}\, 5 \,\textrm{Hz}$ only as higher frequency components will mainly originate from the regions of the simulation with characteristic frequencies smaller than that which correspond to the softened volume. To obtain accurate GW signals of higher frequencies, accretion on to the NS must be modelled. With a hydrodynamically resolved NS one could excite oscillation modes that would result in GWs with frequencies up to kHz, which would be in the range of the LVK network \citep{Schutz2008}.

The long-term evolution of the merger remnant is difficult to predict.
At the point we end the simulation, there is $0.94$\msol of gas gravitationally bound.
If this material were to be accreted onto the NS it could lead to an accretion induced collapse into a BH
 as we approach the maximum mass for a NS depending on the EoS.
The jets have steadily ejected material at a rate of $3\E{-4}M_\odot/\textrm{s}$. 
We cannot predict the long term evolution of this rate, as the launching region enters the unresolved volume at the end of the simulation.
Additionally, the amount of mass ejected by the jet is likely dependant on unresolved or non-modelled physics.
In particular, accurately modelling the accretion onto the NS can dramatically change the outcome.
The accretion would release a great amount of the gravitational energy budget that is contained on the disk that and could contribute to the heating or ejection of the material as well as the release of neutrinos.

Furthermore, nuclear reactions also inject energy into the gas, reducing the amount of gravitationally bound material.
Depending on the amount of mass that these non-modelled effects can unbind, the NS may survive or collapse into a BH.

Despite the limitations of our current simulations, we have shown that a bipolar jet is self-consistently launched. 
The duration of the jetted ejection is difficult to estimate from the period of time computed so far. A rough extrapolation of the mass accretion rate onto the unresolved central compact object implies that the bound mass around the unresolved NS may be swallowed on scales of few thousand seconds. We find that the variability of the jet-luminosity happens down to timescales ${\sim}\, 25\,$s. This is probably an upper bound induced by the moderate numerical resolution of the jets combined with the effective criteria employed to define the matter in the jet. The power at (significantly) smaller variability timescales would be induced by the changes close to the neutron star, which we cannot resolve.

The jet may be sustained for a sizeable fraction of the accretion time. The transient produced by this merger may share a number of qualitative similarities with other jetted ejections resulting from, e.g. the tidal disruption of main-sequence stars by black holes \citep[e.g.][]{Mimica2015a}, with some key differences.
Firstly, due to the compactness of the WD (compared to a $1\,M_\odot$ star), the disruption takes place very close to the NS and, as a result, a large fraction of the disrupted material is still gravitationally bound.
In analogy to jetted tidal disruption events (TDEs), the accretion of most of the WD mass onto the central object on scales of thousands of seconds will likely produce a very luminous transient.
Secondly, the jets launched during this event may asymptotically reach velocities of the order of the escape velocity at the jet basis.

Assuming the NS not to collapse to a BH despite the accretion, we expect asymptotic jet velocities close to that observed in objects such as galactic microquasars \citep[${\gtrsim}\, 0.9 c$, e.g.][]{Mirabel1998},  with typical Lorentz factors of $2$ to $10$. 
As a result, the spectrum of this event could be characterized by an initial soft X-ray or UV flash, that will evolve into progressively lower frequencies.
This would make this event resemble some fast blue optical transients \citep[FBOTs, see e.g.][]{Drout2014,Ho2023,Prentice2018,Coppejans2020}. These transients have also been observed in the outskirts of galaxies, suggesting an origin related to old stellar populations, for which a NS-WD merger would be a good match.
Depending on the properties of the interstellar medium surrounding the NS-WD merger, a bright, weeks-to-months lasting radio signal may be observed \citep[c.f.][]{Mimica2015a}.
The luminosity of this transient could be larger than that of typical kilonovae \citep[${\sim}\, 10^{41}\,\text{erg\,s}^{-1}$][]{Metzger2017}, not only because of the larger mass and velocity of the ejecta, but also because the simulated event does not appear to harbor optimal conditions for the production of r-process yields.
The presence of lanthanides in kilonovae obscures these events due to their large contribution to the opacity.

If, however, the NS eventually does collapse into a BH, significantly higher (ultrarelativistic) jet speeds could be produced, resulting in some long or, possibly, ultra-long GRB-like transient \citep{Thone2011, Levan2014}.

We obtain an average neutrino flux of $2\E{48}\textrm{erg}\,\textrm{s}^{-1}$, which would put this transient one order of magnitude below the fluxes predicted for supernovae. However, we caution that in our simulations the neutrino luminosity is dominated by the unresolved conditions next to the NS.

The magnetically driven jet observed in our simulations is promising for explaining some recently discovered astrophysical transients. Therefore, a better numerical resolution of the material in the jet is desirable which can be readily achieved by changing the refinement criteria applied in our \textsc{arepo} simulations.
Future modeling work on mergers between WDs and NSs includes the addition of nuclear reactions to our simulations \citep[for models including nuclear reactions see][]{Bobrick2022}, paying special attention to the treatment of nuclear burning inside the unresolved regions, and a more accurate modelling of the physics around the NS, ideally including accretion onto the compact object.

\section*{ACKNOWLEDGMENTS}
The authors thank Valeriya Korol, Eva Laplace and Maryam Modjaz for their valuable input.
J.M-F. is fellow of the International Max Planck Research School for Astronomy and Cosmic Physics at Heidelberg (IMPRS-HD) and acknowledges financial support from IMPRS-HD.

This work was supported by the European Research Council (ERC) under the European Union’s Horizon 2020 research and innovation programme under grant agreement No.\ 759253 and 945806, the Klaus Tschira Foundation, and the High Performance and Cloud Computing Group at the Zentrum f{\"u}r Datenverarbeitung of the University of T{\"u}bingen, the state of Baden-W{\"u}rttemberg through bwHPC and the German Research Foundation (DFG) through grant no INST 37/935-1 FUGG.

MAA acknowledges the support through the grant PID2021-127495NB-I00 funded by MCIN/AEI/10.13039/501100011033 and by the European Union, the Astrophysics and High Energy Physics program of the Generalitat Valenciana ASFAE/2022/026 funded by MCIN and the European Union NextGenerationEU (PRTR-C17.I1), and the grant CIPROM/2022/13 from the Generalitat Valenciana.

\bibliographystyle{aa}
\bibliography{references.bib}


\begin{appendix}
\section{MRI\label{sec:appendix1}}
\begin{figure*}
    \includegraphics[width=0.97\textwidth]{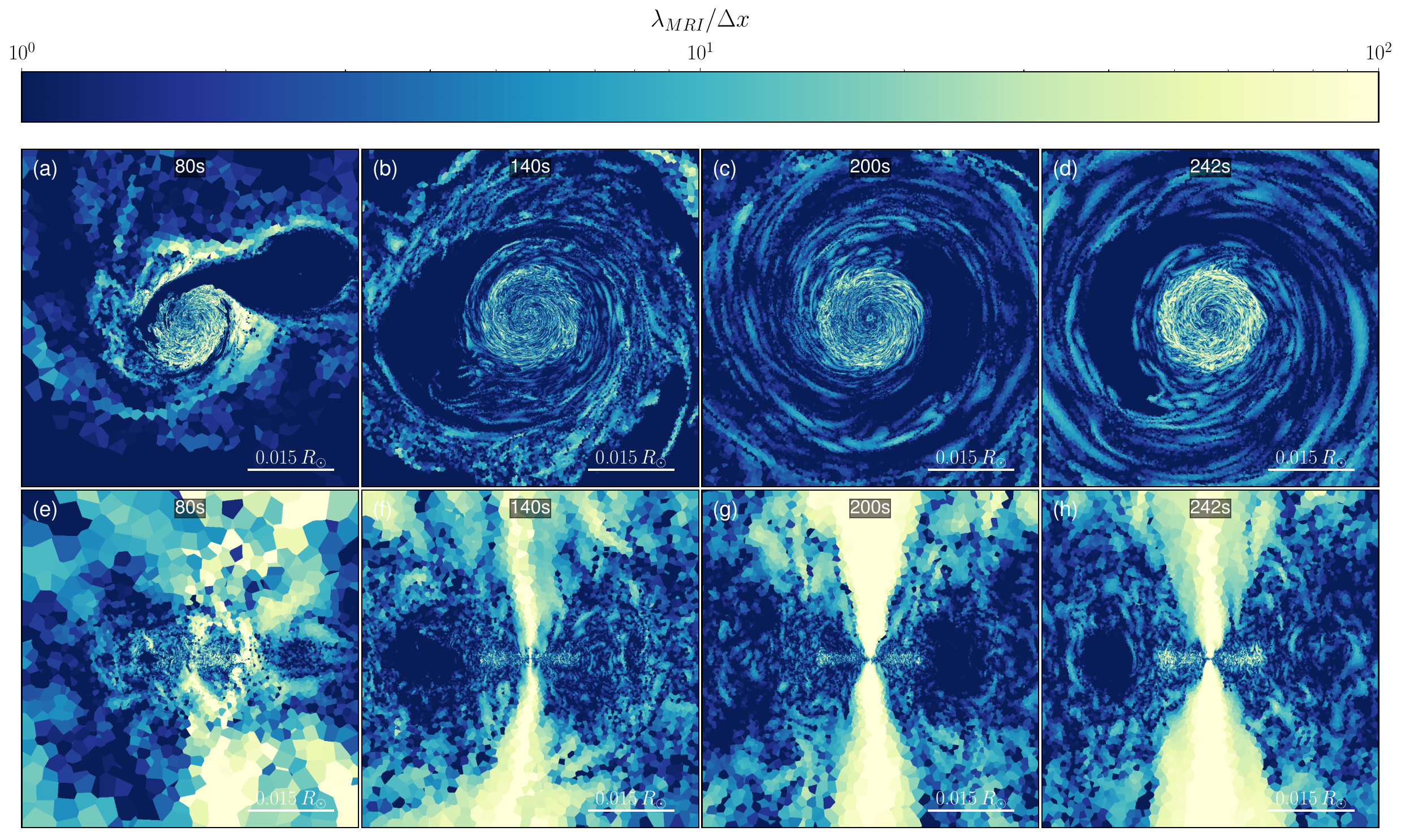}
    \caption{\label{fig:mri}Number of cells per wavelength $\lambda_\textrm{MRI}$ of the fastest growing MRI mode. \emph{Top}: Slices through the orbital plane, centered on the NS.
    \emph{Bottom}: Slices perpendicular to the orbital plane, centered on the NS.}
\end{figure*}

In order to analyze how well we resolve the MRI amplification in our simulation, we have computed the fastest-growing mode following \citet{Rembiasz2016} and compare it to the size of the grid cells in Fig.~\ref{fig:mri}. We find that in the regions closest to the NS, where most of the magnetic field amplification takes place, (see Fig.~\ref{fig:bfld}), the fastest growing mode of the MRI is well resolved.
\end{appendix}

\end{document}